\documentclass[dvips]{article}
\usepackage{supertabular,lscape,epsfig}
\usepackage{amssymb}
\usepackage{amsmath}
\DeclareSymbolFont{ppa}{OT1}{ppl}{m}{it}
\DeclareMathSymbol{\vv}{\mathalpha}{ppa}{'166}

\begin{document}

\newcommand{\dd}{\,{\rm d}}
\newcommand{\ie}{{\it i.e.},\,}
\newcommand{\etal}{{\it et al.\ }}
\newcommand{\eg}{{\it e.g.},\,}
\newcommand{\cf}{{\it cf.\ }}
\newcommand{\vs}{{\it vs.\ }}
\newcommand{\zdot}{\makebox[0pt][l]{.}}
\newcommand{\up}[1]{\ifmmode^{\rm #1}\else$^{\rm #1}$\fi}
\newcommand{\dn}[1]{\ifmmode_{\rm #1}\else$_{\rm #1}$\fi}
\newcommand{\upd}{\up{d}}
\newcommand{\uph}{\up{h}}
\newcommand{\upm}{\up{m}}
\newcommand{\ups}{\up{s}}
\newcommand{\arcd}{\ifmmode^{\circ}\else$^{\circ}$\fi}
\newcommand{\arcm}{\ifmmode{'}\else$'$\fi}
\newcommand{\arcs}{\ifmmode{''}\else$''$\fi}
\newcommand{\MS}{{\rm M}\ifmmode_{\odot}\else$_{\odot}$\fi}
\newcommand{\RS}{{\rm R}\ifmmode_{\odot}\else$_{\odot}$\fi}
\newcommand{\LS}{{\rm L}\ifmmode_{\odot}\else$_{\odot}$\fi}

\newcommand{\Abstract}[2]{{\footnotesize\begin{center}ABSTRACT\end{center}
\vspace{1mm}\par#1\par
\noindent
{~}{\it #2}}}

\newcommand{\TabCap}[2]{\begin{center}\parbox[t]{#1}{\begin{center}
  \small {\spaceskip 2pt plus 1pt minus 1pt T a b l e}
  \refstepcounter{table}\thetable \\[2mm]
  \footnotesize #2 \end{center}}\end{center}}

\newcommand{\TableSep}[2]{\begin{table}[p]\vspace{#1}
\TabCap{#2}\end{table}}

\newcommand{\FigCap}[1]{\footnotesize\par\noindent Fig.\  %
  \refstepcounter{figure}\thefigure. #1\par}

\newcommand{\TableFont}{\footnotesize}
\newcommand{\TableFontIt}{\ttit}
\newcommand{\SetTableFont}[1]{\renewcommand{\TableFont}{#1}}

\newcommand{\MakeTable}[4]{\begin{table}[htb]\TabCap{#2}{#3}
  \begin{center} \TableFont \begin{tabular}{#1} #4
  \end{tabular}\end{center}\end{table}}

\newcommand{\MakeTableSep}[4]{\begin{table}[p]\TabCap{#2}{#3}
  \begin{center} \TableFont \begin{tabular}{#1} #4
  \end{tabular}\end{center}\end{table}}

\newenvironment{references}%
{
\footnotesize \frenchspacing
\renewcommand{\thesection}{}
\renewcommand{\in}{{\rm in }}
\renewcommand{\AA}{Astron.\ Astrophys.}
\newcommand{\AAS}{Astron.~Astrophys.~Suppl.~Ser.}
\newcommand{\ApJ}{Astrophys.\ J.}
\newcommand{\ApJS}{Astrophys.\ J.~Suppl.~Ser.}
\newcommand{\ApJL}{Astrophys.\ J.~Letters}
\newcommand{\AJ}{Astron.\ J.}
\newcommand{\IBVS}{IBVS}
\newcommand{\PASP}{P.A.S.P.}
\newcommand{\Acta}{Acta Astron.}
\newcommand{\MNRAS}{MNRAS}
\renewcommand{\and}{{\rm and }}
\section{{\rm REFERENCES}}
\sloppy \hyphenpenalty10000
\begin{list}{}{\leftmargin1cm\listparindent-1cm
\itemindent\listparindent\parsep0pt\itemsep0pt}}%
{\end{list}\vspace{2mm}}

\def\TYLDA{~}
\newlength{\DW}
\settowidth{\DW}{0}
\newcommand{\dw}{\hspace{\DW}}

\newcommand{\refitem}[5]{\item[]{#1} #2%
\def\REFARG{#3}\ifx\REFARG\TYLDA\else, {\it#3}\fi
\def\REFARG{#4}\ifx\REFARG\TYLDA\else, {\bf#4}\fi
\def\REFARG{#5}\ifx\REFARG\TYLDA\else, {#5}\fi.}

\newcommand{\Section}[1]{\section{#1}}
\newcommand{\Subsection}[1]{\subsection{#1}}
\newcommand{\Acknow}[1]{\par\vspace{5mm}{\bf Acknowledgements.} #1}
\pagestyle{myheadings}

\newfont{\bb}{ptmbi8t at 12pt}
\newcommand{\xrule}{\rule{0pt}{2.5ex}}
\newcommand{\xxrule}{\rule[-1.8ex]{0pt}{4.5ex}}
\def\thefootnote{\fnsymbol{footnote}}
\begin{center}

{\Large\bf
Photometric survey for stellar clusters in the outer part of M33}
\vskip1cm
{
\large K. Zloczewski$^1$, J. Kaluzny$^1$, J. Hartman$^2$}
\vskip3mm
{
        $^1$Nicolaus Copernicus Astronomical Center,\\
        ul. Bartycka 18, 00-716 Warsaw, Poland\\
        e-mail: (kzlocz,jka)@camk.edu.pl\\
        $^2$Harvard-Smithsonian Center for Astrophysics,\\
          60 Garden St., Cambridge, MA 02138, USA\\
        e-mail:  jhartman@cfa.harvard.edu
}
\end{center}

\Abstract{We present a catalog of 4780 extended sources from the outer 
field of M33. The catalog includes 73 previously identified clusters
or planetary nebulae, 1153 likely background galaxies, and 3554 new
candidate stellar clusters. The survey is based on deep ground-based
images obtained with the MegaCam instrument on the CFHT telescope.  We
provide $g'r'i'$ photometry for detected objects as well as estimates
of the FWHM and ellipticity of their profiles. The sample includes 122
new, relatively bright, likely globular clusters. Follow-up
observations of fainter candidates from our list may extend the
faint-end of the observed luminosity function of globular clusters in
M33 by up to 3 magnitudes. The catalog includes several cluster
candidates located in the outskirts of the galaxy. These objects are
promising targets for deep photometry with the $HST$. We present a
color-magnitude diagram for one detected object, showing that it is an
extended and low-surface-brightness old cluster located at an angular
distance of 27 arcmin from the center of M33.}

{catalogs-- galaxies: individual (M33) -- 
galaxies: star clusters}

\section{ Introduction}

M33 is an Scd type spiral galaxy belonging to the Local Group.
Following the pioneering work of Hiltner (1960), there have been
several searches for stellar clusters associated with the galaxy.  A
comprehensive summary and overview of this subject can be found in
Sarajedini \& Mancone (2007, SM hereafter).  In particular, these
authors presented an up-to-date catalog of star clusters and cluster
candidates known in the field of M33.  Over the last decade there has
been a significant effort to search for faint clusters as well as for
clusters located in projection near the central part of the galaxy
using the imaging capability of the $HST$ (Chandar et al. 1999, 2001;
Bedin et al. 2005; Park \& Lee 2007).

To date, the faintest clusters known in M33 have $V\approx 20$ or
$M_{\rm V}\approx -5$ (Park \& Lee 2007, SM). In contrast, the least
luminous globular clusters (GCs) detected in the Milky Way (MW) have
$M_{\rm V}\approx -1$ (Koposov et al. 2007). As pointed out by SM, the
sample of clusters in the outskirts of M33 is incomplete.  There is no
evidence for the existence of dwarf spheroidals (dSphs) or tidal tails
connected with M33 (McConnachie et al. 2006). This is in contrast to
M31 and the Milky Way which both harbor sizable populations of dwarf
galaxies and show extended tidal streams of stars (Ibata et al. 1994,
Ibata et al. 2001, Majewski et al. 2006).  Only recently fourteen
faint satellites of the MW were found in the Sloan Digital Sky Survey
data by looking for spatial over-densities of old, metal-poor stellar
populations (e.g Willman et al. 2002, Koposov et al.  2007). Nine of
them are dSphs, two (Koposov 1 and 2) are GCs, the other three have
properties intermediate between these two classes (Walsh et al. 2008,
Martin et al. 2007, Belokurov et al. 2007).  Sarajedini (2007)
suggests that the significant age range of M33 clusters together with
the lack of dwarf satellites or detectable tidal streams favors the
hypothesis that the galaxy is in fact a member of the M31
'super-halo'.

The above facts together with the availability of new observations
prompted us to survey the area of M33 for new stellar clusters; in
particular for new GCs.  In the following sections we describe the
data that we used, the adopted search method, and the catalog of 4780
extended sources detected in the field of M33. The catalog includes
122 new objects which are likely globular clusters.  The paper concludes
with a short analysis of one particularly interesting new stellar
cluster and with a brief summary.

\section{The data}

The imaging data used in this work were obtained using the Queue Service
Observing mode at the Canada France Hawaii Telescope (CFHT) 3.6 m
telescope within the programme "M33 CFHT Variability Survey" (Hartman et
al. 2006, H06 hereafter).  They were collected on 16 separate nights
between August 2003, and November 2004. The data are available on-line
from {\sl The Canadian Astronomy Data Center}. Images were gathered in
$g^{\prime}$, $r^{\prime}$ and $i^{\prime}$ Sloan filters using the
MegaCam instrument which is a wide-field mosaic imager consisting of 36
2048 $\times$ 4612 pixel CCDs. Exposure times of individual frames
varied between 480 and 600 s (see Sec 2 H06 for details). At the
MegaPrime focus the instrument provides the full field of view amounting
to 1-deg$^2$ at a scale of 0.185 arcsec/pixel. The images cover most of
the area occupied on the sky by M33 (see Fig. 1 in H06). Processing of
the CCD frames was performed as a part of the CFHT Queue Service
Observing Mode using the ELIXIR pipeline.

To facilitate the search for, and photometry of, faint extended sources
we stacked a subset of the best images for 
each filter and CCD (field hereafter) combination. 
The input lists include 6-8 images in $g^{\prime}$, 7-10 images in $r^{\prime}$ and
6-10 images in the $i^{\prime}$ filter. Median seeing values of the selected
images (from all 36 fields) are 0.95, 0.88 and 0.82 arcsec for $g^{\prime}$,
$r^{\prime}$ and $i^{\prime}$ respectively. For each field/filter a "reference image"
was chosen and the remaining individual images were then re-sampled to
its coordinates.  This was done using a modified version of the {\sc ISIS
2.1} image subtraction software (Alard and Lupton 1998, Alard 2000).
The main modification of ISIS involved using a list of relatively
unblended stellar sources with good S/N ratio as input for the {\it
interp} task. The suitable stars were extracted using the {\sc DoPhot}
program (Schechter et al. 1993). In running ISIS we divided each
field into 2-8 overlapping sub-fields. This helped us cope with the
spatial variability of the point spread function (PSF). For each
field and filter combination the master frame was created by stacking
a relevant set of re-sampled images. These master images were then
used in the analysis presented below.

\section{The search procedure}

Huxor et al. (2008) summarize the various strategies that have been
employed to search for extragalactic GCs.  In our survey we used a
method similar to that described by Mochejska et al. (1998). The
method is based on the expectation that the profiles of GCs in M33
should be noticeably more extended than the profiles of stars.  As a
consequence, when a stellar PSF is subtracted from the globular
cluster profile, a 'hole' or 'doughnut-shape' structure will be
visible in the residual image.  GCs in the MW exhibit half-light radii
spanning the range 0.5-25~pc with a median value of 3.6~pc (see Table
1 in Mackey \& van den Bergh 2005).  At the distance of M33, such GCs
would show half-light radii of 0.11-5.3~arcsec, with a median value of
0.77~arcsec.  We note that our method may not be the most effective
technique for identifying very extended globular clusters like those
found by Huxor et al. (2005, 2008) in the halo of M31.

The search for GC candidates was performed for the 30 outer-most of
the 36 observed fields. The inner-most fields, numbered by H06 as
13--15 and 22--24, contain a large number of young clusters, which are
beyond the scope of this paper. We note that a successful search for
clusters inside these crowded fields was conducted recently by Park \& Lee
(2007) using HST/WFPC2 archival images. We employed the {\sc
Daophot/Allstar} (Stetson, 1987) package to calculate the PSF and to
perform profile photometry inside the analyzed sub-fields. We used
this software to produce residual images which are free of all sources
detected by {\sc Daophot} and accepted by {\sc Allstar}. Candidate
extended, non-stellar objects were selected by visual inspection of
the residual images.  Preliminary centroids of these candidates were
determined for each of the three filters.  We then used an iterative
procedure to re-build an image of each object. First, all "stars"
located within 3.5 pixels of the center of each object are added back
into the image.  After this step we define a set of "object pixels" to
consist of all pixels within the above radius that have a signal that
is 4-$\sigma$ above the local sky level.  We then check for pixels
next to the object pixels that are 5-$\sigma$ below sky level.  If
such a dip is associated with a previously subtracted star, then the
star is added back into the image.  This procedure usually ends after
a few iterations. The images from the 3 separate filters are aligned
spatially and the master set of object pixels is formed by their
combination.  Finally, for each filter, the image of an extended
object is reconstructed by assigning original values to all object
pixels from the master set.  Each re-constructed image is then
inspected visually to check for bright field stars affecting the
object profile. If such stars are noted they are subtracted from the
image. The steps involved in the above procedure are illustrated in
Fig. 1.

To derive accurate centroids of the extended objects and to estimate the
FHWM of their profiles we have employed the {\sc SExtractor 2.5.0}
package (Bertin \& Arnouts, 1996). We set the package parameters to
detect all objects having more than 10 pixels with values 3-$\sigma$
above the local sky level. Refined object coordinates were calculated
using the windowing option. For details on the background estimation and
the source selection see the {\sc SExtractor} user manual (Bertin \&
Arnouts, 1996; Holwerda 2005).  The code was applied to frames
containing only reconstructed images of the candidate extended objects.
A total of 5243 objects are detected both by visual inspection of the
images and by {\sc SExtractor}.  Of these, 4715 are detected in
$g^{\prime}$, 4794 are detected in $r^{\prime}$ and 4971 are detected in
$i^{\prime}$. Some objects were counted twice as they are located on the
overlapping parts of some sub-fields. After accounting for these
multiple detections we are left with 4780 unique objects, the majority
of which were detected in all three filters.

\subsection{Photometry and astrometry}

We measured the instrumental magnitudes of the extended objects using
2 circular apertures with radii of $1.0\times FWHM$ and $2.0\times
FWHM$. When available, we used the $FWHM$ value returned by {\sc
SExtractor}. In cases where the source was too faint to pass the shape
measurement threshold in {\sc SExtractor} we used the $FWHM$ value
estimated with the $IRAF/imexam$ task\footnote{IRAF is distributed by
the National Optical Astronomy Observatory, which is operated by the
Association of Universities for Research in Astronomy, Inc., under a
cooperative agreement with the National Science Foundation.}. Note
that for 83 faint sources $imexam$ failed to converge as well;
instrumental magnitudes are unavailable for these objects.  Photometry
was performed using the {\sc Daophot/Phot} routine (Stetson 1987).
Instrumental magnitudes were subsequently transformed to the $g'r'i'$
system of H06 by applying zero-point offsets to our measurements. The
offsets were obtained by calculating the sigma-clipped median
difference between the H06 catalog values and the instrumental
magnitudes of stellar sources obtained by us.  We used stars down to
the catalog magnitudes of 23.5.  Our instrumental magnitudes for stars
were obtained by adding appropriate aperture corrections to profile
photometry extracted with {\sc Daophot/Allstar} (Stetson 1987).
Aperture corrections were derived with the {\sc Daogrow} code (Stetson
1990).  The photometry of stellar sources used in H06 is on an
instrumental system which corresponds approximately to the SDSS
$g'r'i'$ system.  The authors applied an approximate transformation
provided by the {\sc ELIXIR} pipeline which was used for processing
the MegaCam data. While implementing the {\sc ELIXIR} calibration H06
used only the zero-point terms and ignored the color terms of the
transformations.  It is expected that a more accurate calibration of
the stellar photometry used in H06 will be available in the near
future. For the moment the photometry of extended objects presented
here is on the same instrumental system as the catalog of H06. One
may, however, use the transformation provided in Sec. 4.1 of H06 to
get approximate $R$ magnitudes of objects in our list.

The astrometric frame solution was derived for each field using the
equatorial coordinates of stellar sources included in the catalog of
H06. The median difference between matched stars was 0.04 arcsec. The
astrometric system of H06 is tied to the USNO-B1.0 catalog (Monet et
al. 2003), however, for most of MegaCam CCDs the calibration was
obtained indirectly using astrometry from Massey et al. (2006).  We
estimate that the accuracy of the equatorial coordinates obtained for
our objects is about 0.25~arcsec.

\Section{The catalog and selection of GC candidates}

Table 1 lists photometric and astrometric data for a total 4780
non-stellar objects. We show here only a few rows from the catalog which
is available in full on the web\footnote{See AAA archive described at
http://www.astrouw.edu.pl/$^{\sim}$acta/acta.html; the data are
available also at http://www.camk.edu.pl/case/results/index.html.}. The
ID's listed in the first column of Table~1 include the MegaCam field
number (see H06), the number of our subfield and the number of the
object on a given subfield. The equatorial coordinates listed in columns
2-3 are followed by photometry listed in columns 4-15. As described
above we provide aperture magnitudes measured for radii of $1.0\times
FWHM$ and $2.0\times FWHM$. Columns 16 and 17 give $FWHM$ and
ellipticity $ell$ measured in the $g^{\prime}$ filter with {\sc
SExtractor}. The flags returned by {\sc SExtractor} for the
$g^{\prime}$, $r^{\prime}$ and $i^{\prime}$ filters are listed in the
next 3 columns. The flags\footnote{Object's flags meaning: $-$1 FW and
elipticity was not measured, 0 no comment, 1 has neighbors bright and
close or bad pixels, 2 was originally blended with another one, 4 at
least one pixel is saturated, 8 too close to image boundary, 16 aperture
data are incomplete or corrupted, 32 isophotal data are incomplete or
corrupted, 64 a memory overflow occurred during deblending, 99 {\sc
SExtractor} have not detected the object, 128 a memory overflow occurred
during extraction. Positive flags but 99 are returned by {\sc
SExtractor}. Please refer to package manual for details.} returned by
{\sc SExtractor} for the $g^{\prime}$, $r^{\prime}$ and $i^{\prime}$
filters are listed in columns 18-20. Column 21 gives the $FWHM$ value
estimated with $IRAF/imexam$. The last column provides our proposed
classification of the object: $-1$ galaxy (1155 objects); $0$
unclassified (3462 objects); $1$ likely stellar cluster (122 objects);
$2$ an already known high confidence clusters included in the SM catalog
(41 objects). Our classification is based on visual inspection of the
images. Most background galaxies show either structures resembling
spiral arms or an elongated "bulge/bar" surrounded by an ellipsoidal
halo.  Images of most candidate stellar clusters show some "clumpiness"
which presumably reflects the presence of some unresolved but noticeable
stars within the limits of the object.  It should be stressed that the
catalog includes only objects which are unresolved or only partially
resolved into stars on the MegaCam images. We did not attempt to include
resolved objects such as stellar associations or relatively extended
young open clusters.\\

One may expect that the sample of unclassified objects is dominated by
background galaxies. This is particularly true for faint objects which
were difficult to classify by visual inspection. Nonetheless,
detecting and studying the faint population of GCs in M33 would be
worth the effort. One way to distinguish between stellar clusters in
M33 and background galaxies would be to use $HST$ images. However,
only a small fraction of the outer field of the galaxy has been imaged
with either WFPC2 or ACS. Another way to select genuine clusters would
be to measure their radial velocities.\\

The sample of targets for spectroscopic follow-up can be narrowed by
applying two filters. First, one may exclude candidates with
ellipticity $e>0.3$. All known globular clusters in the Local Group
have $e<0.3$. Harris et al. (2006) note also a lack of clusters with
$e>0.3$ in a large sample of NGC~5128 clusters.  Applying the
ellipticity criterion to our list reduces the number of objects of
type 1 and 0 from 3584 to 2544. Out of the 1040 rejected objects, 179 do not have an ellipticity measurement, 840 are of type 0 and have $e \geq 0.3$, and 21 are of type 1 and have $e \geq 0.3$.\\

The second possible filter uses the fact that stellar clusters show
either blue or only moderately red colors.
In Fig. 2 we show a color-color diagram of all objects of type 0 and 1 that have $e<0.3$. The confirmed clusters included in the catalog of 
SM are marked with bold symbols on the diagram. 
As one can see, the confirmed clusters occupy a well defined region in
the color-color space. The area inside the quadrilateral marked in
Fig. 2 contains a total of 758 objects with ellipticity $e<0.3$. Fourty nine of them are classified as likely clusters. Of the remaining 50
objects with such classification, 38 are located outside of the
quadrilateral region and 11 do not have magnitude measurements in all
three bands. The color-magnitude diagrams presented in Figs. 3-4 show
that the sample of objects with $g^{\prime}-r^{\prime}<0.7$ or $r^{\prime}-i^{\prime}<0.5$, which can be
considered candidate stellar clusters, extends about 3 mag below
the faint end of the currently confirmed sample.\\ 

In Figs. 5-7 we present images of 122 objects of type 1 included in
our catalog. They can be compared with images of 41 "high confidence
clusters" from the SM catalogue which are shown in Fig. 8.  As one can
see, many objects from our sample look similar to the already known
clusters. Spectroscopic follow-up of the whole sample of new cluster
candidates would be challenging, but feasible, with currently
available instrumentation.

A total of 45 clusters from the SM catalog are located on images of M33
analyzed in our survey. Table 2 gives their cross-identification with
our numbering system. We have recovered 41 of them. The bright cluster
SM-420 is saturated on our images. The compact cluster SM-419 was
identified on HST/WFPC images by Chandar et al. (2001), on our images
its profile is hardly different from profiles of nearby stars. Moreover,
the object is located only 2.7~arcsec from a star of comparable
magnitude. The clusters SM-034 and SM-426 did not pass our visual
selection.

Ten objects (\#03-1-017, \#03-3-014, \#03-3-023, \#04-6-015,
\#05-6-018, \#16-6-008, \#21-8-004, \#25-4-009, \#26-1-013 and
\#33-4-018) match to planetary nebulae 
candidates listed in Magrini et al. (2001). Only \#33-4-018 was
classified by us as a likely cluster.

\Section{Cluster 34-5-022}
Some of the newly detected cluster candidates can be seen in archival
$HST/ACS$ images. We are planning to discuss these objects further in
a separate contribution.  However, it is possible to obtain resolved
stellar photometry using the MegaCam data for at least one of the new
objects that we have detected: the cluster candidate \#34-5-022. As
there are no HST data for this object we present here a
short discussion of its properties.  As can be seen in Fig. 7
\#34-5-022 is a relatively extended object of low central
concentration and low surface brightness.  Its profile has a FWHM of
4.3~acrcsec, and the half-light radius can be estimated at 1.2~arcsec,
which at the distance of M33 corresponds to 5.5~pc. For comparison, in
the sample of 41 high confidence clusters from the SM list which are
included in our catalog, the most extended object is \#12-1-008 with
FWHM=3.34~arcsec.  From a visual inspection of available images the
total radius of cluster \#34-5-022 can be estimated conservatively at
7.4~arcsec or 36~pc. Several stars located inside the limits of the
object can be resolved on the MegaCam images.  In Fig. 9 we present
color-magnitude diagrams for these stars along with photometry of
nearby stars from the galaxy field. The stars located within the
cluster limits primarily occupy the left side of the red giant branch
defined by the M33 field stars. This indicates that \#34-5-022 is most
likely an old or intermediate age object of metallicity lower than the
average metallicity of the surrounding M33 population.\\ 

We have measured $BV$ magnitudes for \#34-5-022 using public images of
M33 collected with the CFHT telescope and the CFHT12K
mosaic. Specifically we have used the images 2002BF02-674528 and
2002BF02-674529 for $B$ and $V$, respectively.  The zero points of the
photometry were derived using a few nearby stars with $BV$ photometry
published by Massey et al. (2006). For the aperture of radius
4.7~arcsec we obtained for
\#34-5-022 $B=20.00$ and $V=19.38$.  Recent direct determinations of
the distance modulus to M33 cover the range of 24.32-24.92 (Bonanos et
al. 2006).  Adopting $(m-M)_{0}=24.6$ and $A_{\rm V}=0.3$ we obtain an
absolute magnitude of the cluster $M_{\rm V}=-5.2$. This luminosity
places the object in the domain of globular clusters in the $M_{\rm
V}$ versus half-luminosity-radius plane. According to the empirical
formula proposed by Mackey \& van den Bergh (2005), at $M_{\rm
V}=-5.2$ the division between dwarf galaxies and globular clusters
occurs near $R_{h}=45$pc.  We conclude that \#34-5-022 is a very good
candidate for one of the faintest and at the same time one of the most
extended old stellar clusters detected so far in M33. It is located in
the outskirts of the galaxy disk at a projected angular distance
from the center of $d=26.8$~arcmin.  Of a total of 264 high confidence
clusters included in the SM catalog, only 3 objects have larger values of
distance with $27.3<d<28.2$~arcmin. Moreover, Stonkute et al. (2008)
recently reported the detection of an extended cluster, M33-EC1,
located at a projected distance of 48.4~arcmin from the galaxy
center. This cluster as well as cluster \#34-5-022 deserve detailed
photometric study using $HST$ images.

\Section{Discussion}
The presented catalog of extended objects located in the outer field
of M33 contains 122 probable globular clusters.  They are first class
candidates for spectroscopic follow-up.  The catalog also includes a
few hundred fainter objects that are diffuse and roughly
spherical. Studying this sub-sample may allow one to extend the faint
end of the currently known luminosity function of M33 clusters by 2-3
magnitudes.  Our list includes 33 likely clusters (objects of type 1)
which are located at a distance of $25<d<33$~arcmin from the center of
M33. The SM catalog includes only 3 clusters and 1 unclassified object
which are located in this distance range. The clusters located in the
outer part of the galaxy are attractive targets for obtaining deep
color-magnitude diagrams with the $HST$. The diagrams of clusters
projected onto the inner part of the disk are strongly dominated by
field stars (Sarajedini et al. 1998, 2000) making it difficult to
distinguish their red giant branches and nearly impossible to
identify their main sequence stars. This problem should be greatly
reduced for the outer-most clusters. Moreover, studying these objects
will provide insight on the puzzling paucity of blue horizontal branch
clusters in M33.  The color-magnitude diagrams of 3 outer M33 disk
fields presented by (Barker et al. 2007) include stars located about
3.5 mag below the level of the red giant clump.  These diagrams are
based on relatively shallow images obtained with the $HST/ACS$. Hence,
one may expect that deeper exposures of the outer-most clusters should
allow clear identification of the upper main sequence stars in these
objects.

\Acknow{Research of JK and KZ is supported by the  Foundation for the 
Polish Science through grant MISTRZ. 
This research used the facilities of the Canadian Astronomy Data 
Centre operated by the National Research Council of Canada with the 
support of the Canadian Space Agency.
Based on observations made with the NASA/ESA Hubble Space Telescope, 
and obtained from the Hubble Legacy Archive, which is a collaboration between 
the Space Telescope Science Institute (STScI/NASA), the Space Telescope 
European Coordinating Facility (ST-ECF/ESA) 
and the Canadian Astronomy Data Centre (CADC/NRC/CSA).
Based on observations obtained with MegaPrime/MegaCam, a
joint project of CFHT and CEA/DAPNIA, at the Canada-France-Hawaii
Telescope (CFHT) which is operated by the National Research Council (NRC)
of Canada, the Institut National des Science de l'Univers of the Centre
National de la Recherche Scientifique (CNRS) of France, and the University
of Hawaii.
}

\clearpage

\begin{table}[h!]
{\footnotesize
\begin{center}

\begin{tabular}{| c | c c | c c c c c c |}
\hline
ID & ${\alpha}_{2000}$ [$^\circ$] & ${\delta}_{2000}$ [$^\circ$] & $g^{\prime}$ (1)& $g^{\prime}$err (1)& $r^{\prime}$ (1)& $r^{\prime}$err (1)& $i^{\prime}$ (1)& $i^{\prime}$err(1)\\
\hline
01-1-001   & 23.90952  & +31.10895  & 24.051 & 0.069 & 23.258 & 0.034 & 22.676 & 0.033\\
01-1-002   & 23.91051  & +31.12007  & 23.760 & 0.036 & 22.596 & 0.014 & 21.508 & 0.008\\
01-1-003   & 23.91083  & +31.08312  & 99.999 & 9.999 & 99.999 & 9.999 & 99.999 & 9.999\\
01-1-004   & 23.91130  & +31.09520  & 23.292 & 0.026 & 22.160 & 0.010 & 21.368 & 0.008\\
01-1-005   & 23.91165  & +31.05102  & 23.345 & 0.045 & 22.159 & 0.013 & 21.548 & 0.012\\
01-1-006   & 23.91170  & +31.10112  & 23.140 & 0.025 & 22.011 & 0.010 & 20.954 & 0.006\\
\hline
\end{tabular}

\begin{tabular}{|c c c c c c | c c r r r | c | c |}
\hline
$g^{\prime}$ (2)& $g^{\prime}$err (2)& $r^{\prime}$ (2)& $r^{\prime}$err (2)& $i^{\prime}$ (2)& $i^{\prime}$err (2)& FW[$''$] & ell & \multicolumn{3}{c|}{flags} & FWHM[$''$] & type\\
\hline
23.518 &0.084 &23.196 &0.071 &22.609 &0.069 &   1.63  &  0.152 &  0 & 99 &  0 & 1.07  &  0\\
23.177 &0.045 &22.186 &0.020 &21.150 &0.012 &   1.20  &  0.247 &  0 & 99 &  0 & 1.07  &  0\\
99.999 &9.999 &99.999 &9.999 &99.999 &9.999 &  -1.00  & -1.000 & -1 & 99 & 99 & -1.00 &  0\\
22.651 &0.032 &21.869 &0.017 &21.112 &0.014 &   1.26  &  0.258 &  0 &  0 &  0 & 1.12  &  0\\
23.227 &0.090 &21.920 &0.023 &21.281 &0.020 &   1.54  &  0.249 &  0 &  0 &  0 & 1.12  &  0\\
22.336 &0.027 &21.481 &0.013 &20.512 &0.009 &   1.38  &  0.097 &  0 &  0 &  0 & -1.00 &  0\\
\hline
\end{tabular}

\caption{Catalogue of candidate clusters in the outer part of M33. 
}

\end{center}
}
\end{table}

\begin{table}[h!]
{\footnotesize
\begin{center}
\begin{tabular}{| c c | c c | c c | c c |}
\hline
SM  &  ID      & SM  & ID       & SM  & ID       & SM  & ID\\
\hline
003 & 25 8 024 &        035 & 25 3 001 &        327 & 31 6 020 &        436 & 12 2 015 \\
006 & 25 8 015 &        041 & 33 4 016 &        345 & 31 6 012 &        438 & 30 6 013 \\
009 & 25 7 008 &        053 & 33 6 006 &        393 & 04 6 011 &        439 & 12 3 011 \\
010 & 16 4 016 &        090 & 33 3 008 &        401 & 04 6 014 &        441 & 12 6 003 \\
015 & 25 2 017 &        091 & 06 6 015 &        421 & 12 2 005 &        443 & 12 5 006 \\
016 & 34 1 026 &        102 & 33 3 005 &        422 & 12 1 006 &        446 & 12 5 016 \\
023 & 25 2 014 &        115 & 06 6 018 &        425 & 12 2 008 &        447 & 12 7 022 \\
026 & 34 2 006 &        149 & 05 3 005 &        427 & 12 2 010 &        449 & 11 1 024 \\
028 & 07 6 018 &        155 & 32 6 016 &        428 & 12 1 008 &        & \\
029 & 16 6 013 &        194 & 05 6 005 &        429 & 12 2 014 &        & \\
032 & 16 7 016 &        216 & 32 3 019 &        432 & 12 1 011 &        & \\
\hline
\end{tabular}
\caption{
Cross-identification of high confidence clusters from the SM
catalog with objects from our list.
}
\end{center}
}
\end{table}

\begin{figure}[htb]
\centerline{\includegraphics[width=126mm]{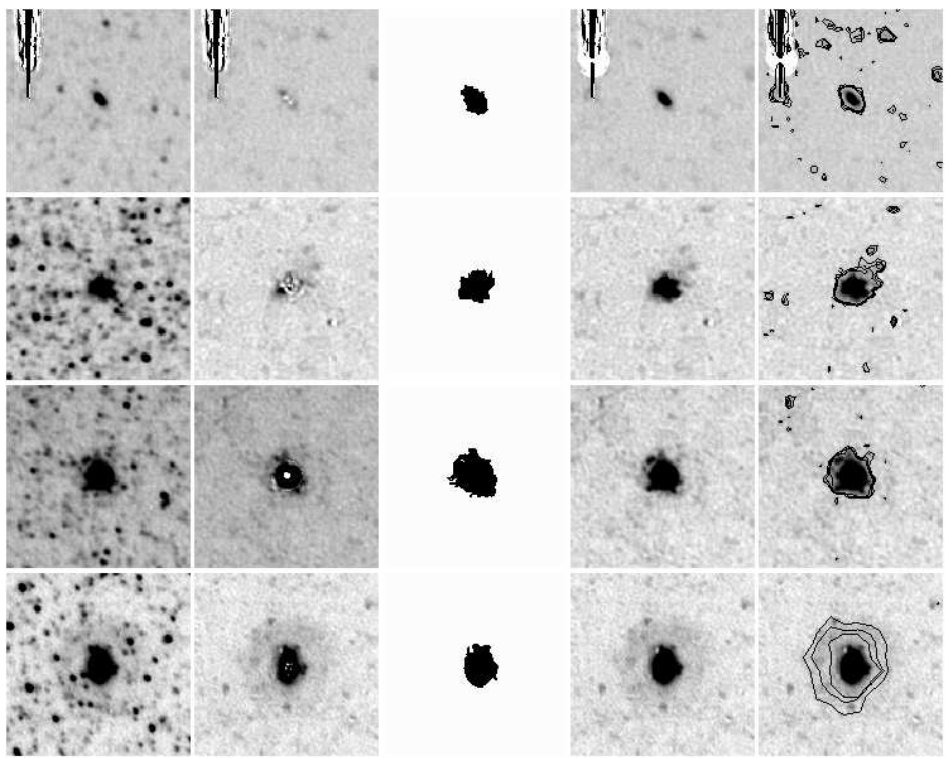}}
\caption{Illustration of a procedure to construct clean
images of extended objects. Rows from top to bottom show: 
a background galaxy (\# 02-01-022), a new cluster candidate
(\# 12-07-007), a known cluster (\# 34-02-006)
and a background galaxy with an extended halo.
Columns from left to right show: input image, star subtracted image,
object pixels  map , cleaned object image and  cleaned object image with  
marked contours of 2-$\sigma$, 3-$\sigma$ and 4-$\sigma$ counts above
sky level. Each stamp has a size of 25$''$ $\times$ 25 $''$.}
\end{figure}

\begin{figure}[htb]
\centerline{\includegraphics[width=126mm]{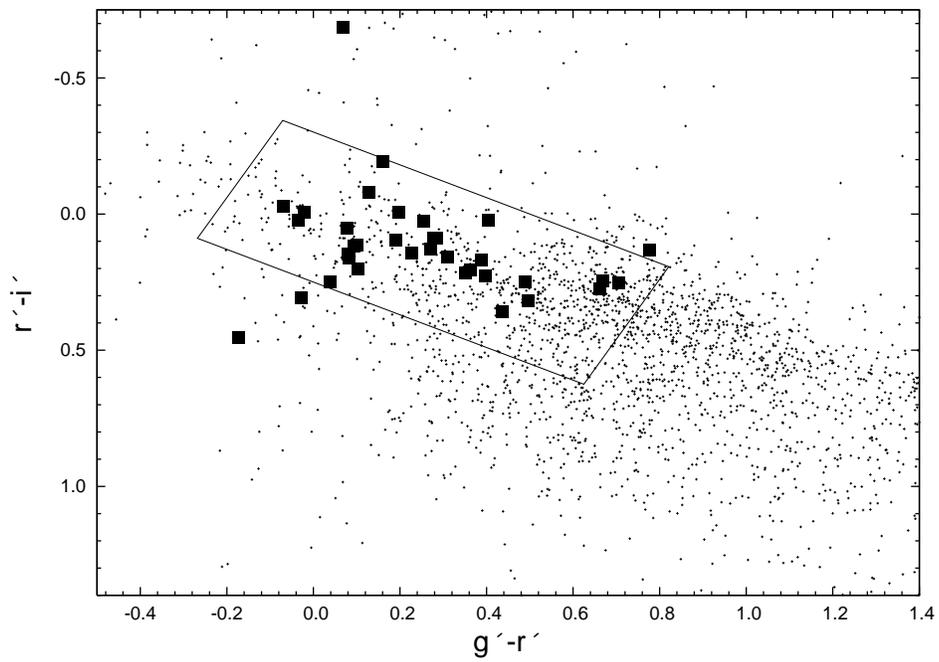}}
\caption{Colour-colour diagram for objects of type 0 or 1 and 
with ellipticity $e<0.3$.  Filled squares are high confidence clusters
from the SM catalog. The parallelogram shows a color selection that
may be applied to choose candidate clusters for spectroscopic
follow-up.}
\end{figure}

\clearpage 

\begin{figure}[htb]
\centerline{\includegraphics[width=126mm]{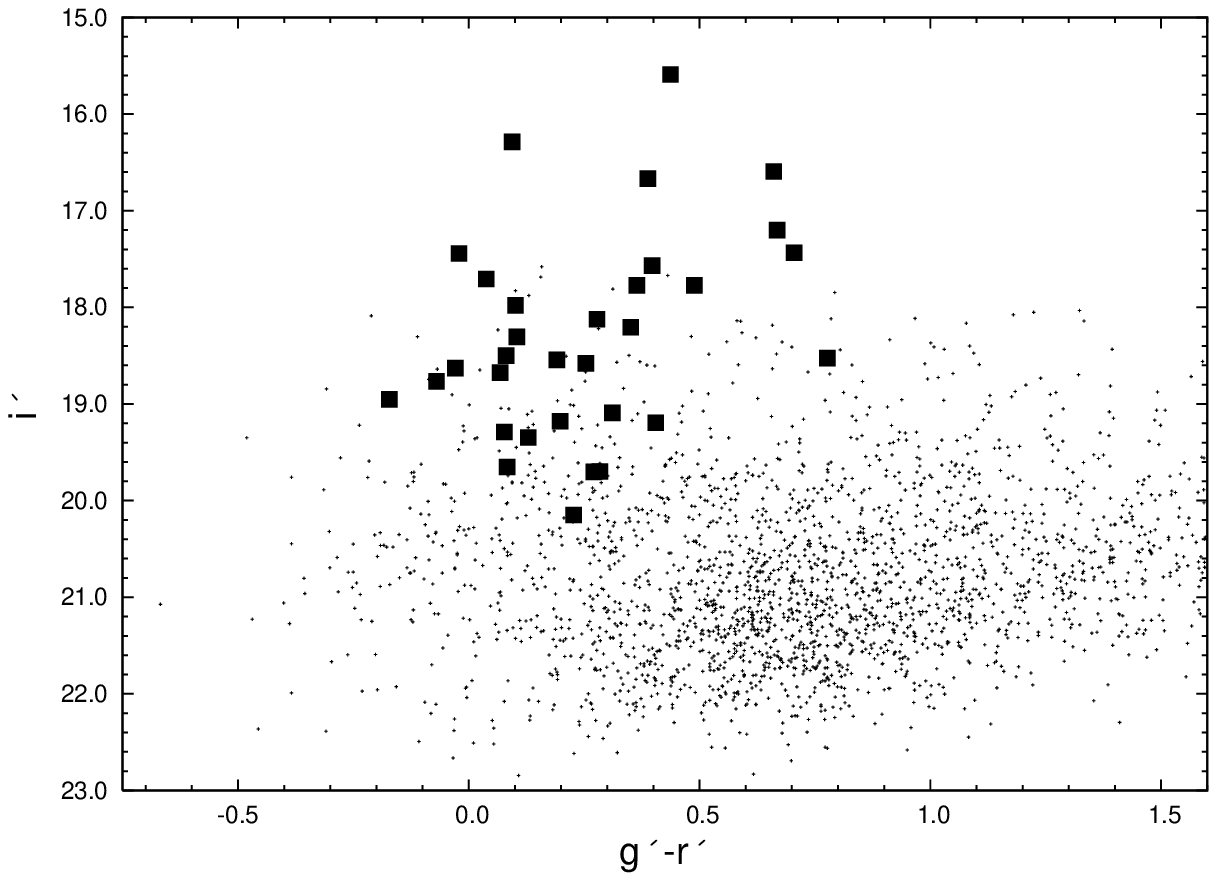}}
\caption{$g^{\prime}-r^{\prime}$ vs.  $i^{\prime}$ diagram 
for objects of type 0 or 1 and $e<0.3$.  
Filled boxes are high confidence clusters from  
the SM catalog.}
\end{figure}

\begin{figure}[htb]
\centerline{\includegraphics[width=126mm]{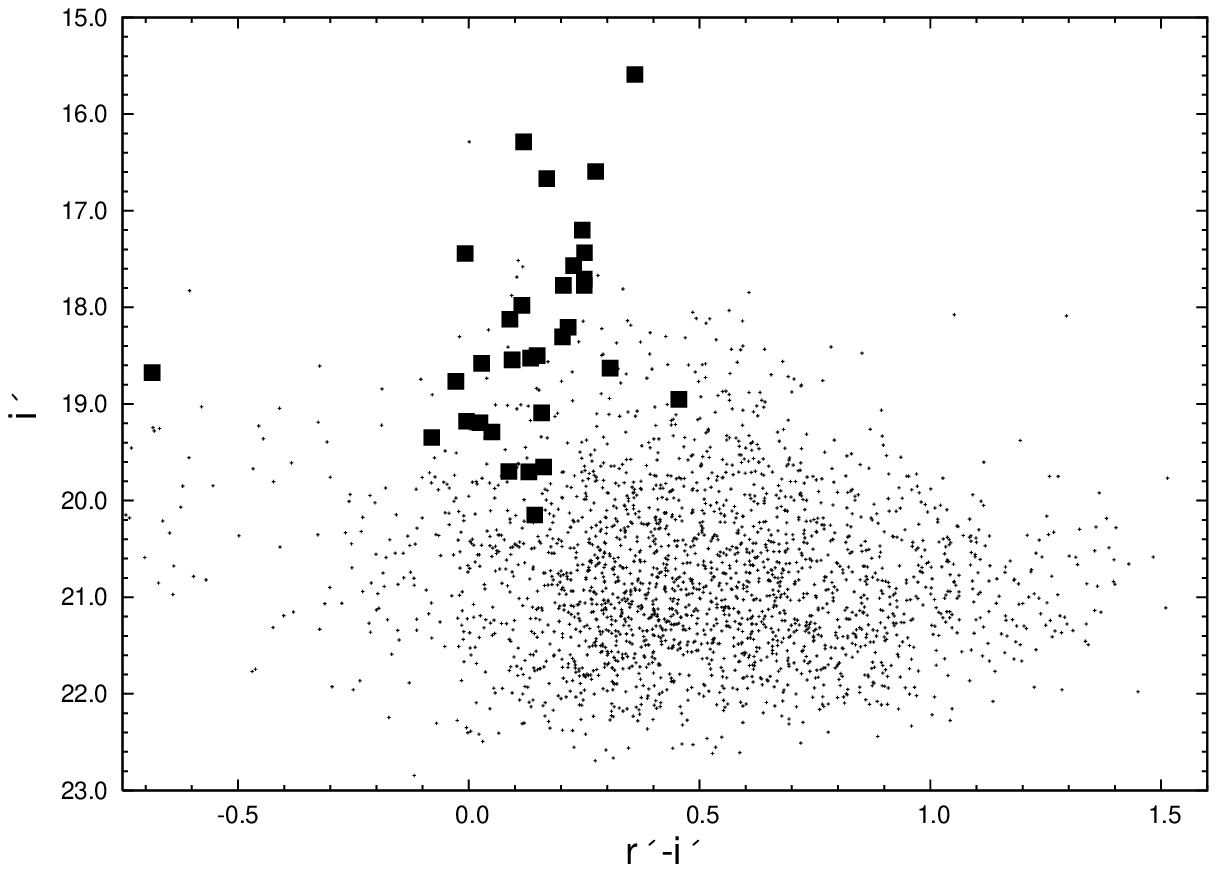}}
\caption{$r^{\prime}-i^{\prime}$ vs. $i^{\prime}$ diagram 
for objects of type 0 or 1 and $e<0.3$.
Filled boxes are high confidence clusters from
the SM catalog.}
\end{figure}

\clearpage

\begin{figure}[htb]
\centerline{\includegraphics[width=166mm]{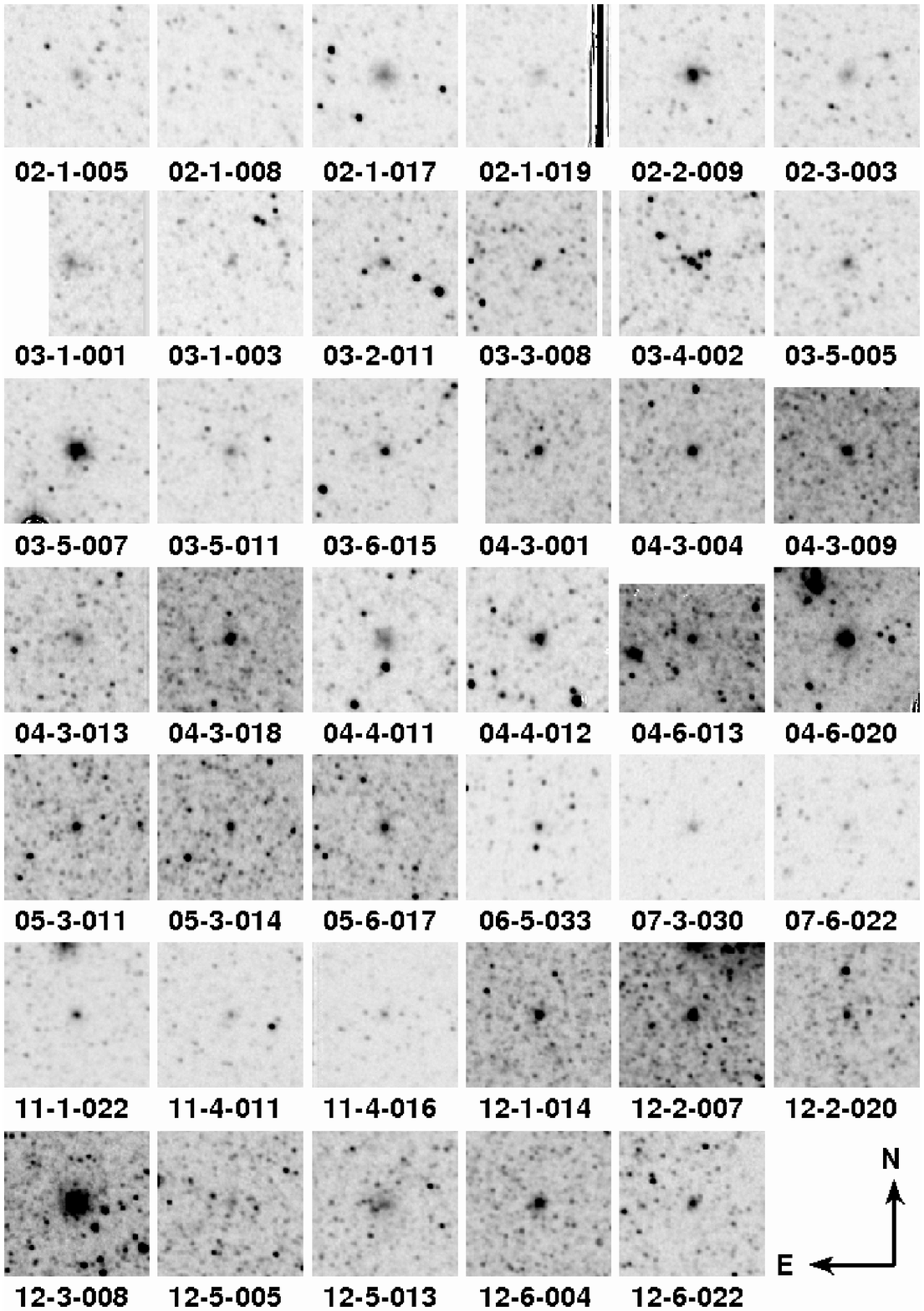}}
\caption{Globular cluster candidates in M33. 
Each stamp has 25 arcsec on a side. North is up and east is to the left.}
\end{figure}

\clearpage

\begin{figure}[htb]
\centerline{\includegraphics[width=166mm]{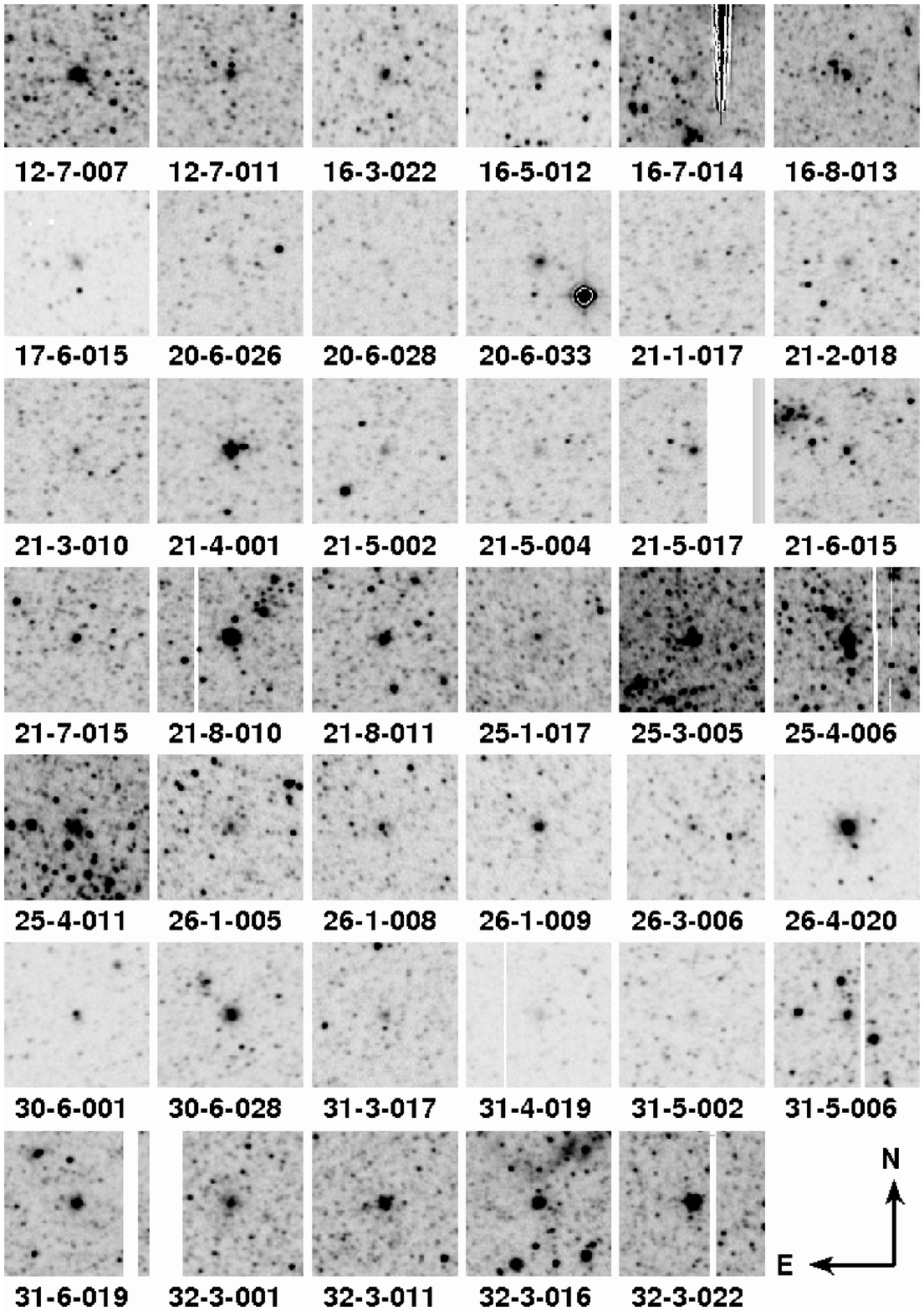}}
\caption{Same as Fig. 5.}
\end{figure}

\clearpage

\begin{figure}[htb]
\centerline{\includegraphics[width=166mm]{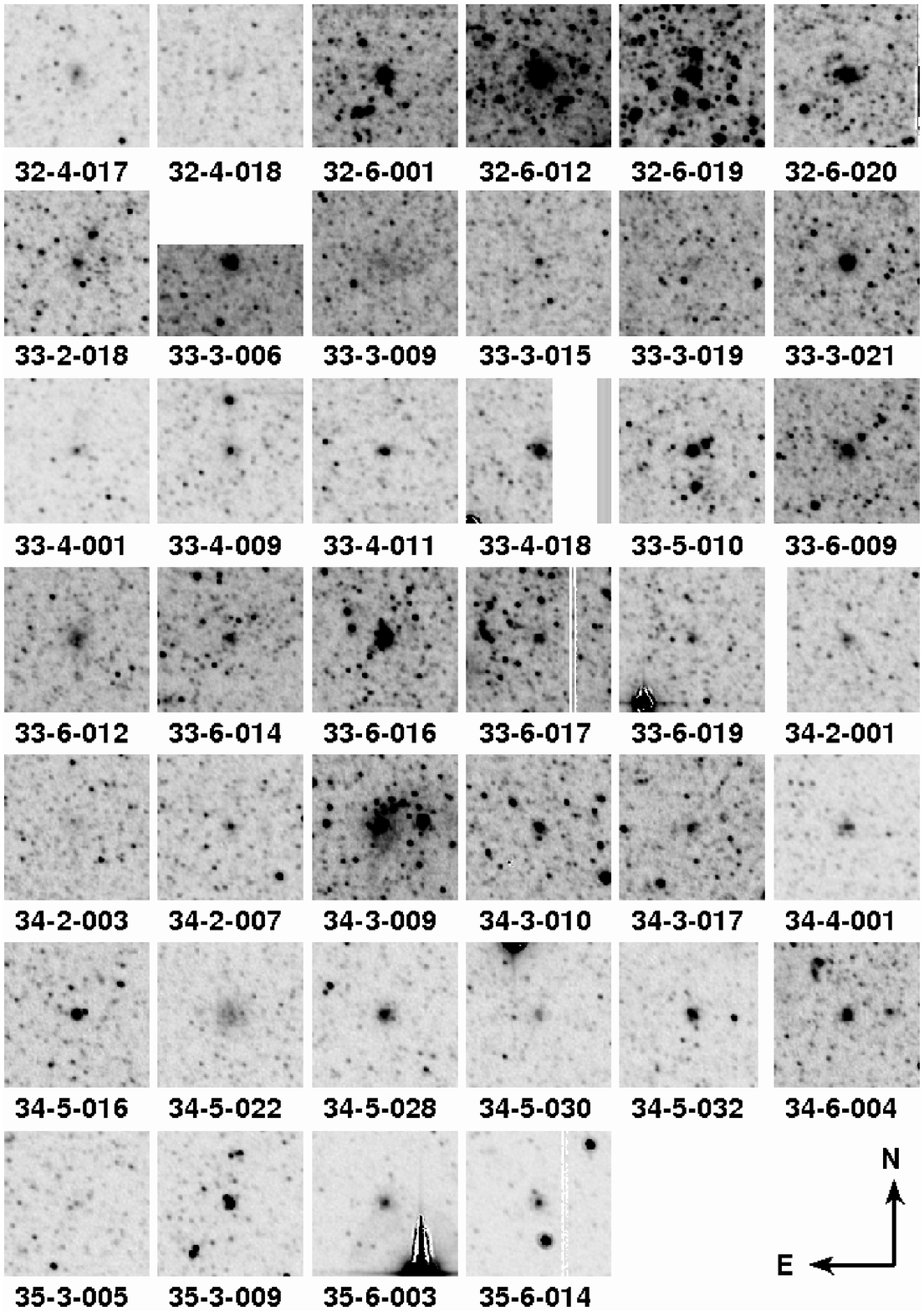}}
\caption{Same as Fig. 5.}
\end{figure}

\clearpage

\begin{figure}[htb]
\centerline{\includegraphics[width=146mm]{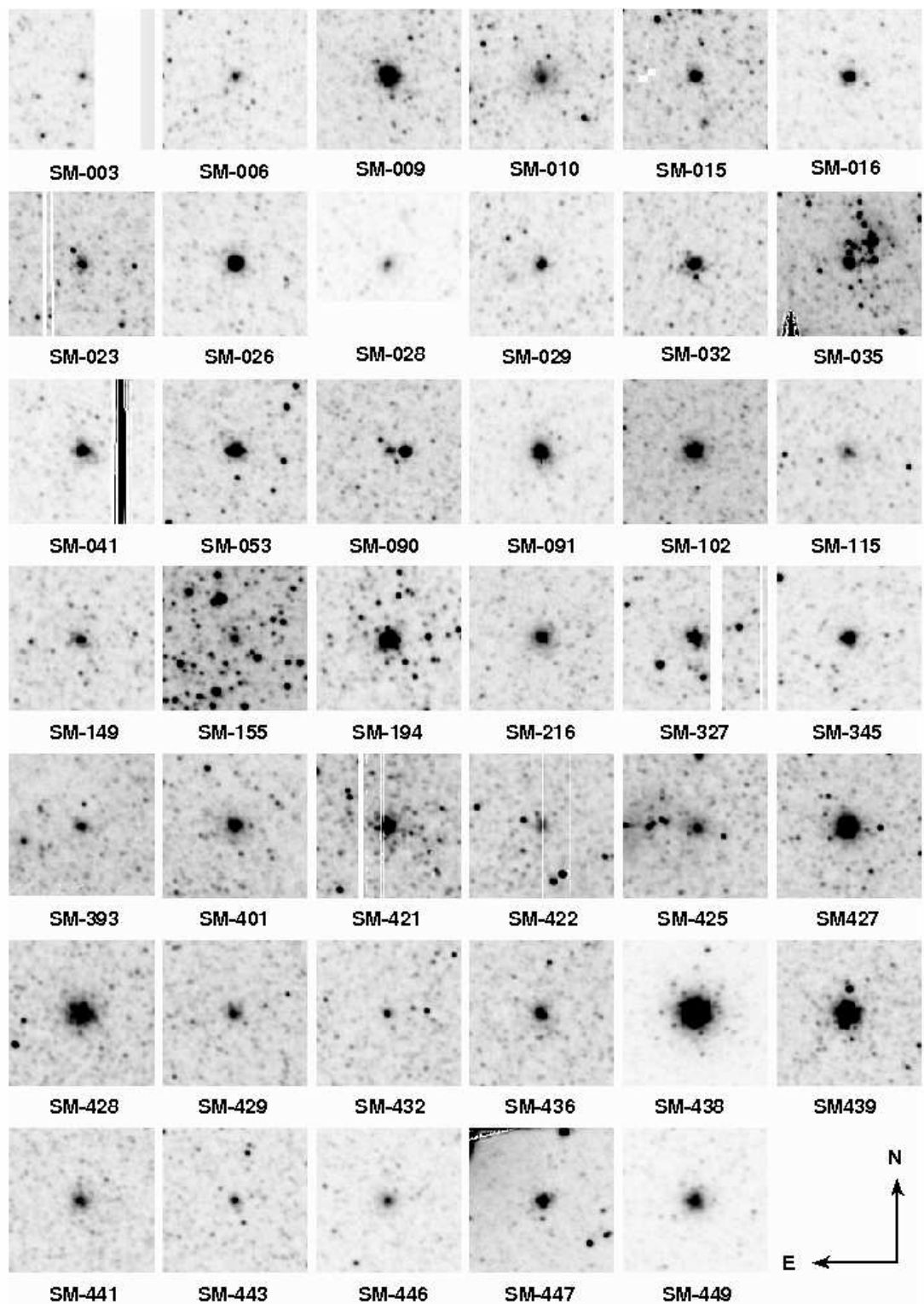}}
\caption{High confidence clusters from the SM catalogue 
included in our list. Each stamp has 25 
arcsec on a side. North is up and east is to the left.}
\end{figure}

\clearpage 

\begin{figure}[htb]
\centerline{\includegraphics[width=126mm]{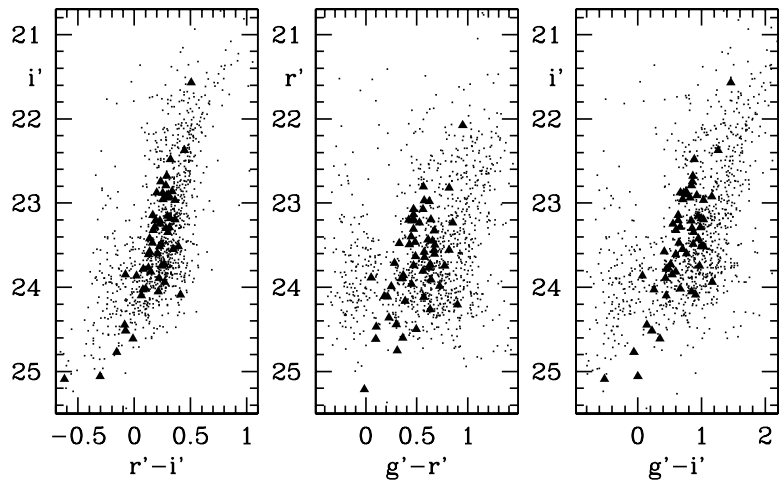}}
\caption{Color-magnitude diagrams for the 
$90\times 90$~arcsec$^{2}$ region of M33
centered on the stellar cluster \#34-5-022. Stars located at
a projected distance $d<4.8$~arcsec from the cluster center
are marked with large symbols.
}
\end{figure}

\end{document}